\documentclass[a4paper,12pt]{article}

\usepackage{epsf}
\usepackage{amsmath}
\usepackage{amssymb}
\usepackage{graphicx}
\usepackage{graphics}
\usepackage{dcolumn}
\usepackage{bm}
\usepackage{color}

\begin{document}
\begin{center}
\Large\bf\boldmath
\vspace*{0.8cm} Constraints on charged Higgs bosons from 
$D^\pm_s\to \mu^\pm\nu$ and $D^\pm_s\to \tau^\pm\nu$
\unboldmath
\end{center}
\vspace{0.6cm}
\begin{center}
A.G. Akeroyd$^a$\footnote{Electronic address: \tt akeroyd@ncu.edu.tw} and F. Mahmoudi$^b$\footnote{Electronic address: 
\tt mahmoudi@in2p3.fr} \\[0.4cm]
\vspace{0.6cm}
{\sl a: Department of Physics, National Central University, Jhongli, Taiwan 320}\\
{\sl b: Laboratoire de Physique Corpusculaire de Clermont-Ferrand (LPC), Universit\'e Blaise Pascal, CNRS/IN2P3, 63177 Aubi\`ere Cedex, France}
\end{center}

\vspace{0.7cm}
\begin{abstract}
\noindent The decays $D^\pm_s\to \mu^\pm\nu_{\mu}$ and 
$D^\pm_s\to \tau^\pm\nu_{\tau}$ have traditionally been used to measure the
$D^\pm_s$ meson decay constant $f_{D_s}$. Recent measurements at  CLEO-c
and the $B$ factories suggest a branching ratio for both decays 
somewhat higher than the Standard Model prediction using 
$f_{D_s}$ from unquenched lattice calculations. 
The charged Higgs boson ($H^{\pm}$) in the Two Higgs Doublet Model (Type II) would
also mediate these decays, but any sizeable contribution 
from $H^{\pm}$ can only suppress the branching ratios and consequently 
is now slightly disfavoured.
It is shown that constraints on the parameters $\tan\beta$ and 
$m_{H^\pm}$ from such decays can be competitive with and complementary to analogous 
constraints derived from the leptonic meson decays $B^\pm\to \tau^\pm\nu_\tau$ and $K^\pm\to \mu^\pm\nu_\mu$, especially if lattice calculations eventually prefer $f_{D_s}< 250$ MeV.\\
\end{abstract}

\vspace{0.3cm}

\section{Introduction}
In the Standard Model (SM) the purely leptonic decays of charged 
pseudoscalar mesons ($P^\pm\to \ell^\pm\nu$) proceed via annihilation in the 
$s$-channel to a $W$ boson. The decay rates are proportional to the 
lepton mass $m_\ell$ which arises from the chirality flip
of the lepton required to conserve angular momentum. Such decays have traditionally been used to measure
the decay constants of the pseudoscalar mesons, and thus provide
an important test of lattice Quantum Chromodynamics (QCD) calculations.
The unprecedented data samples 
provided by the $B$ factories have enabled the 
decay $B^\pm\to \tau^\pm\nu$ to be observed \cite{Ikado:2006un, Adachi:2008ch, Aubert:2007bx,Aubert:2007xj}
despite its relatively small
branching ratio (BR) and challenging signature. The decays
$D^\pm\to \mu^\pm\nu$ \cite{Artuso:2005ym,:2008sq}, $D^\pm_s\to \mu^\pm\nu$ 
\cite{Pedlar:2007za,Artuso:2007zg,Alexander:2009ux} and $D^\pm_s\to \tau^\pm\nu$ 
\cite{Pedlar:2007za,Alexander:2009ux,:2007zm,Onyisi:2009th}
have been measured with $\sim 10\%$ precision at CLEO-c, and the recently commenced 
BES-III experiment will provide improved measurements \cite{Li:2008wv,Asner:2008nq}.
Moreover, the decay $D^\pm_s\to \mu^\pm\nu$ has been measured at the $B$ factories 
\cite{:2007ws,Aubert:2006sd} with a precision slightly less than that of CLEO-c.

These purely leptonic decays are sensitive to charged Higgs boson ($H^\pm$) at the {\sl tree level} 
and thus provide valuable probes of such particles which are complementary to constraints
provided by loop-induced decays (e.g., $b\to s\gamma$). 
Importantly, in supersymmetric (SUSY) models the loop-induced decays are
particularly sensitive to the sparticle spectrum and the
assumptions made for the SUSY breaking sector, and thus the purely leptonic
decays offer a more model-independent probe of parameters in the Higgs sector.  
The measurement of $B^\pm\to \tau^\pm\nu$ \cite{Ikado:2006un,Aubert:2007bx},
although in rough agreement with the SM prediction, provides important
constraints on the mass and couplings of $H^\pm$
from the Two Higgs Doublet Model (Type~II) \cite{Hou:1992sy},
with some dependence on SUSY parameters \cite{Akeroyd:2003zr,Itoh:2004ye,Chen:2006nua,Nierste:2008qe,Isidori:2006pk,Barenboim:2007sk} 
which enters at higher orders in perturbation theory. Improved measurements of
$B^\pm\to \tau^\pm\nu$ (and first observation of $B^\pm\to \mu^\pm\nu$) are thus certainly desirable and 
such decays play a prominent role in the physics case for a high luminosity flavour factory
\cite{Yamauchi:2002ru,Bigi:2004kn,Hashimoto:2004sm,Akeroyd:2004mj,Browder:2004wu,Hewett:2004tv,Bona:2007qt,Browder:2007gg,Browder:2008em}.

The decay rates of $D^\pm \to \mu^\pm\nu$ and $D^\pm_s \to \mu^\pm\nu/\tau^\pm\nu$ 
have traditionally been considered as a robust test of lattice QCD calculations of the decay constants
$f_D$ and $f_{D_s}$ because significant New Physics effects were thought to be unlikely.
However, the effect of $H^\pm$ on $D^\pm_s \to \mu^\pm\nu/\tau^\pm\nu$
was pointed out to be non-negligible in 
\cite{Hou:1992sy,Hewett:1995aw,Hewett:1994hh} and a numerical study \cite{Akeroyd:2003jb,Akeroyd:2007eh} showed that the magnitude of the
$H^\pm$ contribution can be comparable to both the CLEO-c error and the lattice QCD error for the decay constants.
SUSY particles from models with $R$-parity violating interactions
can also contribute to both $D^\pm \to \mu^\pm\nu$ and $D^\pm_s \to \mu^\pm\nu/\tau^\pm\nu$
\cite{Akeroyd:2002pi}. Recently, there has been growing interest in the 
effect of New Physics particles on $D^\pm_s \to \mu^\pm\nu/\tau^\pm\nu$
\cite{Dobrescu:2008er} because the
current world average of the experimental measurements of their branching ratios is
somewhat higher than the SM rate using unquenched lattice QCD calculations of $f_{D_s}$
\cite{Rosner:2008yu, Amsler:2008zzb}. Such a discrepancy cannot be explained by $H^\pm$
of the popular Two Higgs Doublet Model (Type~II) since any sizeable contribution 
(which arises from a relatively large strange quark Yukawa coupling) can only
suppress the decay rate \cite{Hou:1992sy,Hewett:1995aw,Akeroyd:2003jb,Akeroyd:2007eh}.
In this paper we consider in detail the effect of $H^\pm$ on $D^\pm_s \to \mu^\pm\nu/\tau^\pm\nu$
using the program SuperIso \cite{Mahmoudi:2007vz,Mahmoudi:2008tp}. It is shown that constraints derived on the 
plane $[\tan\beta,m_{H^\pm}]$ can be competitive with 
the constraints obtained from the analogous leptonic decays 
$B^\pm\to \tau^\pm\nu$ and $K^\pm\to \mu^\pm\nu$. In particular, we compare the
potential of  $D^\pm_s \to \mu^\pm\nu/\tau^\pm\nu$ and $K^\pm\to \mu^\pm\nu$
\cite{Antonelli:2008jg,Isidori:2008qp,Eriksson:2008cx} to probe the plane $[\tan\beta,m_{H^\pm}]$.

The paper is organized as follows: in section 2 the theoretical formalism for the 
decays $D^\pm_s \to \mu^\pm\nu/\tau^\pm\nu$ and $K^\pm\to \mu^\pm\nu$ is introduced;
numerical results are contained in section 3 with conclusions in section 4.

\section{Effect of $H^\pm$ on leptonic decays of $D^\pm_s$ and $K^\pm$}

In this section we review the theoretical formalism for the effect of 
$H^\pm$ on $D^\pm_s \to \mu^\pm\nu$, $D^\pm_s \to \tau^\pm\nu$ and $K^\pm\to \mu^\pm\nu$,
and summarize the experimental situation.

\subsection{The decays $D^\pm_s \to \mu^\pm\nu$ and $D^\pm_s \to \tau^\pm\nu$}

In the SM the purely leptonic decays  $D^\pm_s\to \ell^\pm\nu_\ell$ proceed via annihilation of the 
heavy meson into $W^*$. Singly charged Higgs bosons,
which arise in any extension of the SM with
at least two $SU(2)_L\times U(1)_Y$ Higgs doublets with hypercharge
$Y=1$, would also contribute to these decays \cite{Hou:1992sy}.
The tree--level partial width is given by (where $\ell=e,\mu$ or $\tau$):
\begin{equation}
\Gamma(D^\pm_s\to \ell^\pm\nu_\ell) = \frac{G_F^2}{8\pi} f_{D_s}^2 m_{\ell}^2 M_{D_s}
\left(1-\frac{m_{\ell}^2}{M_{D_s}^2}\right)^2 \left|V_{cs}\right|^2 r_s\;, \label{equ_rate}
\end{equation}
where in the Two Higgs Doublet Model (2HDM) of Type II 
\cite{Hou:1992sy,Hewett:1995aw,Akeroyd:2007eh,Dobrescu:2008er,Rosner:2008yu}:
\begin{equation}
r_s=\left[1+\left(\frac{1}{m_c+m_s}\right)\left(\frac{M_{D_s}}{m_{H^+}}\right)^2\left(m_c-
\frac{m_s\tan^2\beta}{1+\epsilon_0 \tan\beta}\right) \right]^2  \;.
\label{rs}
\end{equation}
Here $m_c$ and $m_s$ are the masses of the charm and strange quarks respectively, $m_{H^+}$ is the
mass of the charged Higgs boson, $M_{D_s}$ is the mass of the $D^\pm_s$ meson, $\tan\beta=v_2/v_1$
where $v_1$ and $v_2$ are the vacuum expectation values of the two scalar doublets, $V_{cs}$ is
a Cabibbo-Kobayashi-Maskawa (CKM) matrix element, $m_{\ell}$ is the lepton mass and $G_F$ is the Fermi constant.   
In the 2HDM (Type~II) 
each fermion receives mass from one vacuum expectation value
($v_1$ or $v_2$) at tree level
and consequently $\epsilon_0=0$. 
In the 2HDM (Type III) each fermion receives 
a mass from both $v_1$ and $v_2$ and the term $\epsilon_0\ne 0$ must be included. The Yukawa couplings 
of the Minimal Supersymmetric SM (MSSM) 
take the form of a 2HDM (Type~II) at tree level, but at higher orders the structure  
becomes of the type 2HDM (Type III) in which $\epsilon_0$ is a function of SUSY parameters
\cite{Hall:1993gn,Hempfling:1993kv,Carena:1994bv,Blazek:1995nv} and $|\epsilon_0|$ can reach values of order 0.01.
 
The above formula Eq.~(\ref{rs}) 
was derived in \cite{Hou:1992sy} for $\epsilon_0=0$ and neglecting the term $m_c$ in 
$(m_c-m_s\tan^2\beta)$. This
$m_c$ term originates from the charm quark Yukawa coupling and increases 
$r_s$ (and was included in \cite{Hewett:1995aw}) but its magnitude is numerically tiny 
and can be neglected given the current experimental and theoretical errors. 
The term $m_s\tan^2\beta$, which 
originates from the strange quark Yukawa coupling, can give rise to a 
non-negligible suppression of $r_s$ for large values of $\tan\beta$. Both \cite{Hou:1992sy} and \cite{Hewett:1995aw}
neglected $m_s$ when it appears as ($m_c+m_s$) and this correction was first included in \cite{Akeroyd:2007eh}. 
The above complete expression for the case of $\epsilon_0=0$ is taken from 
\cite{Dobrescu:2008er,Rosner:2008yu}. The $\epsilon_0$ correction can be non-negligible for large
$\tan\beta$ and should be included in a dedicated study of $r_s$ in the context of the MSSM.
The parameterization of the $\epsilon_0$ correction in Eq.~(\ref{rs}) is valid in MSSM models with minimal
flavour violation. We neglect to write explicitly the analogous $\epsilon_0$ correction for the $m_c$ term since
this correction does not have the $\tan\beta$ enhancement factor, and the $m_c$ 
term is negligible anyway. A first quantitative study of the magnitude of $r_s$ (setting $\epsilon_0=0$) was performed in 
\cite{Akeroyd:2003jb,Akeroyd:2007eh} for the $m_s\tan^2\beta$ term only
and it was shown that $r_s$ could be suppressed by an 
amount comparable to both the CLEO-c experimental precision and the theoretical 
error in the lattice calculations of $f_{D_s}$. Note that the magnitude of the $H^\pm$
contribution depends on the ratio of quark masses $m_s/(m_c+m_s)$, and an analogous uncertainty
is not present for the $H^\pm$ contribution to the decay $B^\pm\to  \ell^\pm\nu$.

There are various unquenched lattice calculations of $f_{D_s}$
and the current situation is summarized in \cite{Rosner:2008yu} and updated
in \cite{Alexander:2009ux, Gamiz:2008iv}.
The value with the smallest quoted error is $f_{D_s}= 241 \pm 3 \mbox{ MeV}$
\cite{Follana:2007uv} which employs staggered fermions. Another calculation
with staggered fermions gives $f_{D_s}=249\pm 11$ MeV \cite{Gamiz:2008iv, Aubin:2005ar}. 
Other calculations use different lattice techniques and have larger errors
than that of \cite{Follana:2007uv}, although usually with a central value 
less than 250 MeV (e.g., $248\pm 3\pm 8$ MeV \cite{Blossier:2008dj}). 
On the experimental side, $D^\pm_s \to \tau^\pm\nu$ has been measured at CLEO-c 
for two decay modes of $\tau$ \cite{Alexander:2009ux,Onyisi:2009th}
and $D^\pm_s \to \mu^\pm\nu$ has been measured at CLEO-c \cite{Alexander:2009ux}, 
BELLE \cite{:2007ws} and BABAR  \cite{Aubert:2006sd}.
Taking an average of these measurements (excluding the BABAR result which
has an additional error from normalizing to the branching ratio of $D^\pm_s\to \phi^0\pi^\pm$)
results in an unexpectedly high value of $f_{D_s}=261\pm 7$ MeV (derived in \cite{Alexander:2009ux} and is dominated by the average of the three CLEO-c measurements in \cite{Alexander:2009ux} and \cite{Onyisi:2009th}). A previous average gave $f_{D_s}=273\pm 10$ MeV \cite{Rosner:2008yu,Amsler:2008zzb}.
The sizeable difference between this world average measurement of $f_{D_s}$
and the lattice QCD value with smallest error ($f_{D_s}= 241 \pm 3 \mbox{ MeV}$
\cite{Follana:2007uv})
has brought attention to the effect of New Physics particles on these decays
\cite{Dobrescu:2008er}. As discussed above, $H^\pm$ in the 2HDM (Model~II) can only
give a sizeable {\it suppression} of the branching ratio for 
$D^\pm_s \to \mu^\pm\nu,\tau^\pm\nu$, and not a sizeable enhancement.
Candidate models which can enhance the BRs sufficiently
are $R$-parity violating models \cite{Akeroyd:2002pi,Dobrescu:2008er,Kundu:2008ui}
or models with leptoquarks \cite{Dobrescu:2008er,Benbrik:2008ik}.
We note that the scenario of a charged Higgs boson from an unconventional 2HDM with a large 
charm quark Yukawa coupling \cite{Dobrescu:2008er} (which is never possible in standard 
2HDMs in which the top and charm quark receive mass from the same vacuum expectation value)
was also available to accommodate the
data for BR($D^\pm_s\to \mu^\pm\nu/\tau^\pm\nu$). However, such a $H^\pm$ would
enhance BR($D^\pm\to \mu^\pm\nu$) by the same amount \cite{Kronfeld:2008gu} and
now appears to be disfavoured by the final CLEO-c measurement of 
BR($D^\pm\to \mu^\pm\nu$) and $f_D$ \cite{:2008sq}, which is in good agreement with the lattice
calculation of $f_D$ in \cite{Follana:2007uv}. 

In \cite{Narison:2008bc} two lattice QCD calculations \cite{Follana:2007uv, Aubin:2005ar} 
of $f_{D_s}$ were combined with the value obtained from QCD spectral sum rules in order to give an 
average value of $f_{D_s}$. Moreover, an analogous average of calculations of $f_D$ and $f_{D_s}/f_D$ 
were multiplied together to give an independent evaluation of $f_{D_s}$. 
The above procedure gives the value $f_{D_s}=240\pm 7$ MeV 
(i.e. very similar to $f_{D_s}= 241 \pm 3 \mbox{ MeV}$ of \cite{Follana:2007uv}).
The consequences of relaxing the assumption of unitarity of the CKM matrix
when extracting the measurement of $f_{D_s}$ from BR($D^\pm_s\to \mu^\pm\nu/\tau^\pm\nu$)
were also discussed. In the models we study in this work (2HDM and MSSM) the CKM matrix is unitary and
$|V_{cs}|\sim |V_{ud}|$ with very small error. As discussed in \cite{Narison:2008bc},
this scenario leads to the largest discrepancy between the experimental and theoretical values of 
$f_{D_s}$.

In order to make conclusive statements about the possibility of New Physics enhancing
$D^\pm_s\to \mu^\pm\nu/\tau^\pm\nu$ it is evident that a detailed discussion of
the errors for the value of $f_{D_s}$ evaluated in \cite{Follana:2007uv} is required, 
as well as a consensus in the lattice community concerning the
magnitude of the error. Our view is that the current situation of the world average
measurement of $f_{D_s}$ being higher than most lattice calculations of
$f_{D_s}$ is provocative and certainly merits attention because more
precise measurements of $D^\pm_s\to \mu^\pm\nu/\tau^\pm\nu$ will be possible
at the recently commenced BES-III experiment \cite{Li:2008wv}. In addition, 
high luminosity flavour factories are also being discussed \cite{Yamauchi:2002ru}
which have the option of running at the charm threshold \cite{Asner:2007kw}
with data samples even larger than those
anticipated at BES-III.
In this work our aim is not to explain the currently higher than expected 
BR($D^\pm_s\to \mu^\pm\nu/\tau^\pm\nu$) by invoking specific
New Physics models. Indeed, one cannot predict
the preferred values for the lattice calculations of $f_{D_s}$ although one
expects a gradual decrease of the error. The current
discrepancy between the experimental and lattice QCD values for $f_{D_s}$
might persist, worsen or disappear entirely. Instead, we wish to 
emphasize that BR($D^\pm_s\to \mu^\pm\nu/\tau^\pm\nu$) should be considered 
seriously as a process which potentially offers constraints on models containing 
a $H^\pm$ with the popular structure of the 2HDM (Type II), which includes the MSSM.  

In \cite{Rosner:2008yu,Amsler:2008zzb} the SM prediction 
for BR($D^\pm_s\to \mu^\pm\nu_{\mu}/\tau^\pm\nu_{\tau}$) using the 
lattice calculation with smallest error ($f_{D_s}= 241 \pm 3 \mbox{ MeV}$ \cite{Follana:2007uv}) 
was compared with the experimental world average at the time for $f_{D_s}$ ($273\pm 10$ MeV).
It was concluded that any $H^\pm$ contribution is disfavoured at more than $3\sigma$. 
Such a result shows the potentially important role of the decays 
$D^\pm_s\to \mu^\pm\nu/\tau^\pm\nu$ as a probe of $H^\pm$.
However, it is also clear that constraints on $H^{\pm}$ 
derived from these decays are strongly dependent on the central value and error for the 
lattice calculation  of $f_{D_s}$ and so should be interpreted with caution at present. 
In this work we aim to quantify the constraints on the plane $[\tan\beta,m_{H^\pm}]$ from 
$D^\pm_s\to \mu^\pm\nu_{\mu}/\tau^\pm\nu_{\tau}$ with various assumptions for the
central value and error of the lattice QCD calculation of $f_{D_s}$, and compare the
 constraints to those obtained 
from other decays which are sensitive to $H^\pm$.

\subsection{The decay $K^\pm\to\mu^\pm\nu$}
A decay which has many similarities with  $D^\pm_s \to \mu^\pm \nu$ and $D^\pm_s \to \tau^\pm \nu$
is $K^\pm \to \mu^\pm \nu$. This decay can also be mediated by $H^\pm$ of the
2HDM (Type~II) at tree level
\cite{Hou:1992sy} with a scale factor similar to that in Eq.~(\ref{rs}).
In order to reduce the theoretical uncertainties from the decay constant $f_K$, the ratio of partial widths
is usually considered:
\begin{eqnarray}
\dfrac{\Gamma(K^\pm \rightarrow \mu^\pm \nu_\mu)}{\Gamma(\pi^\pm \rightarrow \mu^\pm \nu_\mu)}&=& 
\left|\frac{V_{us}}{V_{ud}} \right|^2\frac{f^2_K m_K}
{f^2_\pi m_\pi}\left(\frac{1-m^2_\ell/m_K^2}{1-m^2_\ell/m_\pi^2}\right)^2 \nonumber \\
&& \times \left(1-\frac{m^2_{K^+}}{m^2_{H^+}}\left(1 - \frac{m_d}{m_s}\right)\frac{\tan^2\beta}{1+\epsilon_0\tan\beta}\right)^2 \left(1+\delta_{\rm em}\right).
\end{eqnarray}
Here $\delta_{\rm em} = 0.0070 \pm 0.0035$ is a long distance electromagnetic correction factor. As suggested in \cite{Antonelli:2008jg}, we study instead the quantity 
\begin{equation}
R_{\ell 23}\equiv\left| \frac{V_{us}(K_{\ell 2})}{V_{us}(K_{\ell 3})} \times \frac{V_{ud}(0^+ \to 0^+)}{V_{ud}(\pi_{\ell 2})} \right|.
\end{equation}
Here $V_{us}(K_{\ell i})$ refers to $V_{us}$ as measured in leptonic decay of $K^\pm$ with $i$ particles in the final state (two leptons and a number of pions), and similarly for $V_{ud}$. $V_{ud}(0^+ \to 0^+)$ 
denotes $V_{ud}$ measured from nuclear beta decay.
In the SM $R_{\ell 23}=1$, while the contribution from $H^\pm$ in the MSSM attains the simple form
(which can be compared to  Eq.~(\ref{rs})):
\begin{equation}
R_{\ell 23}=\left|1-\frac{m^2_{K^+}}{m^2_{H^+}}\left(1 - \frac{m_d}{m_s}\right)\frac{\tan^2\beta}{1+\epsilon_0\tan\beta}\right|.
\end{equation}
Using $m_d/m_s=1/20$ \cite{Amsler:2008zzb}, the MSSM prediction can be directly compared to the experimental value \cite{Antonelli:2008jg}
\begin{equation}
R_{\ell 23}=1.004\pm 0.007.
\label{Rl23}
\end{equation}
In the extraction of this value, the ratio $f_K/f_\pi$ has been fixed to the value $f_K/f_\pi=1.189\pm 0.007$ 
obtained from lattice QCD using staggered quarks \cite{Follana:2007uv}. It should be noted that the uncertainty
 thus obtained for $R_{\ell 23}$ is most probably overly optimistic. Indeed, many approaches exist to determine
 $f_K/f_\pi$, and some reservations remain about staggered fermions despite their impressive success in 
agreeing with 
the experimental values of various ``gold-plated'' observables in flavour physics (e.g., contrasting 
opinions concerning their theoretical viability are expressed in \cite{Creutz:2007rk} and \cite{Kronfeld:2007ek}).
If, for example, 
the value $f_K/f_\pi = 1.205\pm 0.018$ (obtained using the domain wall formulation \cite{Allton:2007hx}) is
 used instead, then $R_{\ell 23}$ provides no constraints on the plane $[\tan\beta, m_{H^\pm}]$
\cite{Eriksson:2008cx}. A recent analysis in \cite{Carena:2008ue} used $f_K/f_\pi = 1.19\pm 0.015$
and also found very weak constraints. We therefore 
stress that the constraints obtained from Eq.~(\ref{Rl23}), although certainly valuable due to their
sensitivity to $H^\pm$ at tree level, 
should serve only as an indication of the potential
of $K^\pm$ decays to probe the plane $[\tan\beta, m_{H^\pm}]$.  
It is our aim to compare such constraints in this optimum scenario for $R_{\ell 23}$
(i.e. using the lattice calculations of \cite{Follana:2007uv})
with those derived from $D^\pm_s\to \mu^\pm\nu/\tau^\pm\nu$. We note that $R_{\ell 23}$ also requires 
lattice QCD input for the semi-leptonic kaon decay form factor $f_+$, while
the decays $D^\pm_s\to \mu^\pm\nu/\tau^\pm\nu$ rely on lattice QCD only for $f_{D_s}$. The direct measurement of $V_{us}$ (relevant for $R_{l23}$) is much more precise than that for $V_{cs}$ (relevant for
$D^\pm_s\to \mu^\pm\nu/\tau^\pm\nu$). However, $V_{cs}$ can be taken as a well-measured parameter by assuming a unitary CKM matrix, which is the case for the models we study.

\section{Numerical Analysis}

The branching ratios for $D^\pm_s \to \tau^\pm \nu_\tau$ and $D^\pm_s \to \mu^\pm \nu_\mu$ have been implemented in 
SuperIso \cite{Mahmoudi:2007vz,Mahmoudi:2008tp}
following Eqs.~(\ref{equ_rate}) and (\ref{rs}). The $\epsilon_0$ correction for the second quark generation 
has also been added to take into account the SUSY effect. This effect has been neglected in all previous studies of 
BR($D^\pm_s \to \tau^\pm \nu_\tau$) and BR($D^\pm_s \to \mu^\pm \nu_\mu$).

To illustrate the constraining power of these observables, we consider the MSSM with minimal flavour violation (MFV), 
and in particular the Non-Universal Higgs Mass model (NUHM). The NUHM
assumes SUSY breaking mediated by gravity and is characterized by a set of universal parameters at 
the GUT scale $\lbrace m_0$, $m_{1/2}$, $A_0$, $\tan\beta, \mu, M_A \rbrace$. The first four parameters in this set 
are the same as in CMSSM (or mSUGRA), but the GUT scale mass parameter universality is relaxed for the Higgs sector 
leading to two additional parameters, $\mu$ and $M_A$, which implies in particular that the charged Higgs mass can be 
considered as a free parameter. This additional freedom makes this model attractive for studying the constraints on the
Higgs sector.

To investigate the NUHM parameter space we generate 50,000 random points scanning over the ranges 
$m_0 \in [50,2000]$ GeV, $m_{1/2} \in [50,2000]$ GeV, $A_0 \in [-2000,2000]$ GeV, $\mu \in [-2000,2000]$ GeV, 
$m_A \in [5,600]$ GeV and $\tan\beta \in [1,60]$. For each point we calculate the spectrum of SUSY particle masses and 
couplings using SOFTSUSY 2.0.18 \cite{Allanach:2001kg} and we compute the branching fractions 
$\rm{BR}(D^\pm_s \to \tau^\pm \nu_\tau)$, $\rm{BR}(D^\pm_s \to \mu^\pm \nu_\mu)$ and 
$\rm{BR}(K^\pm \to \mu^\pm \nu_\mu)$ using SuperIso v2.4 
\cite{Mahmoudi:2008tp}. The obtained values are then compared to the experimentally allowed 
intervals while taking uncertainties in consideration.

The PDG08 \cite{Amsler:2008zzb} combined experimental results for the branching fractions of 
$D^\pm_s \to \tau^\pm \nu$ and $D^\pm_s \to \mu^\pm \nu$ are:
\begin{equation}
\rm{BR}(D^\pm_s \to \tau^\pm \nu)= (6.6 \pm 0.6) \times 10^{-2} \;,
\label{Dstaunu_pdg}
\end{equation}
\begin{equation}
\rm{BR}(D^\pm_s \to \mu^\pm \nu)= (6.2 \pm 0.6) \times 10^{-3} \;.
\label{Dsmunu_pdg}
\end{equation}

The final CLEO-c results using 600 pb$^{-1}$ have been recently released \cite{Alexander:2009ux,Onyisi:2009th}
and give smaller central values and errors:
\begin{equation}
\rm{BR}(D^\pm_s \to \mu^\pm \nu)= (5.65 \pm 0.45 \pm 0.17) \times 10^{-3} \;,
\label{Dsmunu_cleo}
\end{equation}
\begin{equation}
\rm{BR}(D^\pm_s \to \tau^\pm \nu)= (5.62 \pm 0.41 \pm 0.16) \times 10^{-2} \;.
\label{Dstaunu_cleo}
\end{equation}
For BR($D^\pm_s \to \tau^\pm \nu$) the average in Eq.~(\ref{Dstaunu_pdg}) is from the
measurements of \cite{Pedlar:2007za,Heister:2002fp,Abbiendi:2001nb,Acciarri:1996bv}.
In order to have an updated world average we replace the previous CLEO-c measurement of  \cite{Pedlar:2007za}
by the new measurement \cite{Alexander:2009ux,Onyisi:2009th} to obtain:
\begin{equation}
\rm{BR}(D^\pm_s \to \tau^\pm \nu)= (5.7 \pm 0.4) \times 10^{-2} \;.
\label{Dstaunu_exp}
\end{equation}

For BR($D^\pm_s \to \mu^\pm \nu$) the constraints on the plane $[\tan\beta,m_{H^\pm}$] turn out to be weaker
than those from BR($D^\pm_s \to \tau^\pm \nu$). Therefore for illustration and comparison
we only average over the CLEO-c \cite{Alexander:2009ux} and BELLE \cite{:2007ws}
measurements of BR($D^\pm_s \to \mu^\pm \nu$) (as suggested in \cite{Rosner:2008yu,Amsler:2008zzb}) 
which are the only measurements without an additional error from 
normalizing to the branching ratio of $D^\pm_s\to \phi^0\pi^\pm$.
Our average is:
\begin{equation}
\rm{BR}(D^\pm_s \to \mu^\pm \nu)= (5.8 \pm 0.4) \times 10^{-3} \;.
\label{Dsmunu_exp_our}
\end{equation}
In our numerical analysis we employ the lattice calculation of the $D^\pm_s$ decay constant with smallest error
\cite{Follana:2007uv}
in order to give the SM prediction for the branching ratios:
\begin{eqnarray}
f_{D_s}= 241 \pm 3 \mbox{ MeV} \;, \\
\rm{BR}(D^\pm_s \to \tau^\pm \nu)=(4.82 \pm 0.14) \times 10^{-2}\;, \\
\rm{BR}(D^\pm_s \to \mu^\pm \nu)=(4.98 \pm 0.15) \times 10^{-3}\;.
\label{staggered-fds}
\end{eqnarray}%
We will comment qualitatively on the case of using a different value with a larger error.\\
\newpage
\noindent We use the most up-to-date PDG08 values for the quark masses \cite{Amsler:2008zzb}:
\begin{equation}
m_s = 104^{+26}_{-34} \text{ MeV} \qquad m_c = 1.27^{+0.07}_{-0.11} \text{ GeV}\;.
\end{equation}
To take into account this uncertainty we vary the ratio $m_s/m_c$ in the interval [0.04,0.12] with $m_s/m_c=0.08$ being the central value.

An estimation of the total theoretical error in the decay rates (Eq.~\ref{equ_rate})
yields about $3\%$ relative uncertainty. The dominant theoretical error is from
the lattice evaluation of $f_{D_s}$, while the error from $V_{cs}$ is very small 
if CKM unitarity is assumed
(which is the case in the models which we study). The errors from SUSY and other parameters in 
Eq.~(\ref{equ_rate}) yield less than $1\%$ relative uncertainty.
In order to undertake a slightly conservative 
approach we consider $4\%$ relative theoretical uncertainty in our numerical analysis. 
Eqs. (\ref{Dstaunu_exp}) and (\ref{Dsmunu_exp_our}) lead to the following allowed 
intervals at 95\% C.L. for $\rm{BR}(D^\pm_s \to \tau^\pm \nu)$:
\begin{equation}
4.8 \times 10^{-2} < \rm{BR}(D^\pm_s \to \tau^\pm \nu) < 6.6 \times 10^{-2} \;,
\label{Dstaunu_interval}
\end{equation}
and for $\rm{BR}(D^\pm_s \to \mu^\pm \nu)$
\begin{equation}
4.9 \times 10^{-3} < \rm{BR}(D^\pm_s \to \mu^\pm \nu) < 6.7 \times 10^{-3}\;,
\label{Dsmunu_interval}
\end{equation}
in which both experimental and theoretical errors are included, 
as in the analysis of \cite{Eriksson:2008cx}.

\begin{figure}[!p]
\begin{center}
\includegraphics[width=8.0cm]{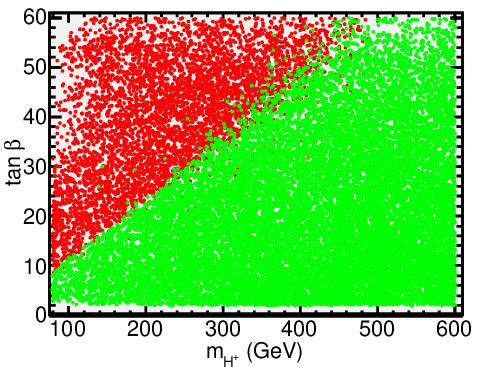}
\includegraphics[width=8.0cm]{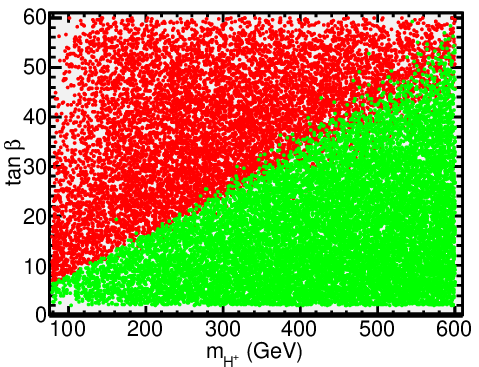}
\includegraphics[width=8.0cm]{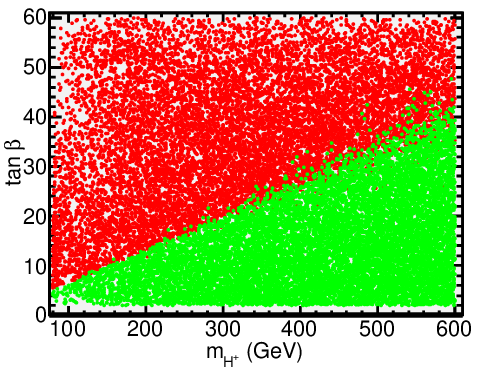}
\caption{Constraints on the plane $[\tan\beta,m_{H^\pm}]$
from the decay $D^\pm_s \to \tau^\pm \nu$ for $m_s/m_c=0.04, 0.08$ and $0.12$ from top to bottom.
Red points are excluded at $95\%$ C.L.}
\label{Dstaunu_fig}
\end{center}
\end{figure}

Fig.~\ref{Dstaunu_fig} illustrates the obtained constraints
by a projection of the six dimensional NUHM parameter space 
onto the plane $[\tan\beta,m_{H^\pm}]$. The allowed points are displayed in the foreground in green while the 
points excluded by $\rm{BR}(D^\pm_s \to \tau^\pm \nu)$ at 95\% C.L are displayed in the background in red. 
The results for three choices of $m_s/m_c$ are presented. A large part of the parameter plane is excluded even for the 
pessimistic case of $m_s/m_c=0.04$. 
The region where red and green points overlap is caused by the theoretical uncertainty (mainly 
arising from $f_{D_s}$) and from the effect of the $\epsilon_0$ term which may take either sign and depends on 
all six NUHM parameters.

\begin{figure}[!p]
\begin{center}
\includegraphics[width=8.0cm]{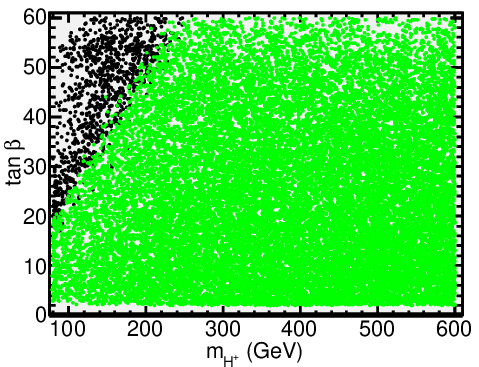}
\includegraphics[width=8.0cm]{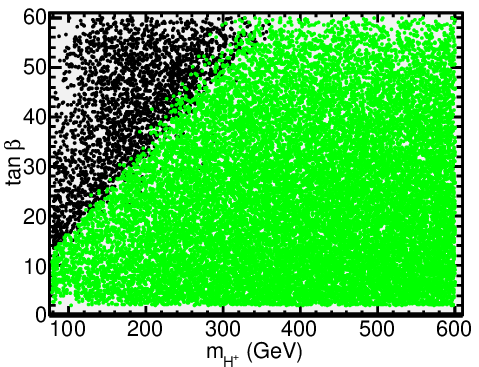}
\includegraphics[width=8.0cm]{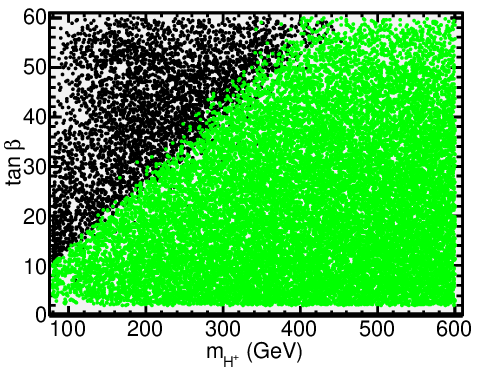}
\caption{Constraints on the plane $[\tan\beta,m_{H^\pm}]$ from the decay
$D^\pm_s \to \mu^\pm \nu$ for $m_s/m_c=0.04, 0.08$ and $0.12$ from top to bottom.
Black points are excluded at $95\%$ C.L.}
\label{Dsmunu_fig}
\end{center}
\end{figure}%

\begin{figure}[!t]
\begin{center}
\includegraphics[width=8.0cm]{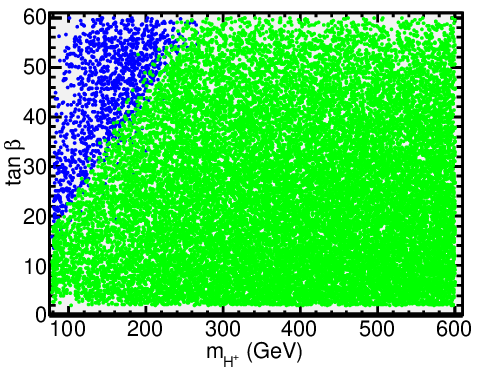}
\caption{Constraints on the plane $[\tan\beta,m_{H^\pm}]$ from the observable
$R_{\ell 23}$ (which involves $K^\pm \to \mu^\pm \nu$/$\pi^\pm \to \mu^\pm \nu$ decays).
Blue points are excluded at $95\%$ C.L.}
\label{Kmunu_fig}
\end{center}
\end{figure}%

In Fig.~\ref{Dsmunu_fig} the points excluded by $\rm{BR}(D^\pm_s \to \mu^\pm \nu)$ are shown in black for 
the same three choices of $m_s/m_c$. The obtained constraints are less severe in this case since the predicted 
SM value is closer to the central experimental value for 
$\rm{BR}(D^\pm_s \to \mu^\pm \nu)$ than for $\rm{BR}(D^\pm_s \to \tau^\pm \nu)$.

In order to compare our results with the constraints from $K^\pm \to \mu^\pm \nu$, we present in Fig.~\ref{Kmunu_fig}
the points excluded at 95\% C.L. by the $R_{\ell 23}$ observable using the range given by Eq.~(\ref{Rl23})
 with no additional error, 
which is probably overly optimistic as discussed in \cite{Eriksson:2008cx} and in section 2.2.
It is clear that $\rm{BR}(D^\pm_s \to \tau^\pm \nu)$ is
more powerful at constraining the plane $[\tan\beta,m_{H^\pm}]$ than $R_{\ell 23}$ even for the pessimistic
case of $m_s/m_c=0.04$ (likewise for $\rm{BR}(D^\pm_s \to \mu^\pm \nu)$ if $m_s/m_c>0.04$).
Although $R_{\ell 23}$ has less uncertainty from quark mass parameters 
than BR($D^\pm_s \to \mu^\pm \nu/\tau^\pm\nu$) (i.e., the theoretical
uncertainty from $m_d/m_s$ is less than that from $m_s/m_c$) we stress that
the pessimistic case of $m_s/m_c=0.04$ already provides competitive constraints on the plane
$[\tan\beta,m_{H^\pm}]$. The lattice techniques discussed in \cite{Davies:2008hs} promise  
precise calculations of $m_c$ and $m_s/m_c$, which would further sharpen potential 
constraints from BR($D^\pm_s \to \mu^\pm \nu/\tau^\pm\nu$).
We note that at a higher confidence level (e.g., 99.7\% C.L.) the constraints on the plane 
$[\tan\beta,m_{H^\pm}]$ from BR($D^\pm_s \to \mu^\pm \nu/\tau^\pm\nu$) are weaker than 
those from $R_{\ell 23}$ because the latter is measured more precisely.

As pointed out in \cite{Antonelli:2008jg},
one of the virtues of $R_{\ell 23}$ is the fact that it can probe a region of 
plane $[\tan\beta,m_{H^\pm}]$ which cannot be probed by the decay $B^\pm\to \tau^\pm\nu$
(corresponding to the case of the $H^\pm$ induced amplitude for $B^\pm\to \tau^\pm\nu$ being 
similar in size to that of the SM contribution but with opposite sign).
We stress that $\rm{BR}(D^\pm_s \to \tau^\pm \nu)$ can exclude this parameter space with greater significance than
$R_{\ell 23}$, using the lattice calculations of \cite{Follana:2007uv} for the SM prediction.
Therefore  both $D^\pm_s \to \tau^\pm \nu$ and $D^\pm_s \to \mu^\pm \nu$ offer
constraints competitive with those of $R_{\ell 23}$ and complementary to those of $B^\pm\to \tau^\pm\nu$.
Moreover, we note that the experimental prospects are more favourable for 
$\rm{BR}(D^\pm_s \to \mu^\pm \nu)$ and $\rm{BR}(D^\pm_s \to \tau^\pm \nu)$ than for $R_{\ell 23}$.
The recently commenced BES-III experiment \cite{Li:2008wv} aims to reduce the error 
of both $\rm{BR}(D^\pm_s \to \mu^\pm \nu)$ and $\rm{BR}(D^\pm_s \to \tau^\pm \nu)$ to the level of
a few percent (i.e. two to three times smaller than the current error)
while a similar improvement for the precision of $R_{\ell 23}$ seems unlikely in the same time scale.

All the above numerical analysis has been performed with $f_{D_s}= 241 \pm 3$ MeV, and
we now comment on the scenario of using a higher value of $f_{D_s}$ with larger errors. 
Repeating the numerical analysis with $f_{D_s}=248\pm 3\pm 8$ MeV \cite{Blossier:2008dj}
results in almost no constraints on the plane $[\tan\beta,m_{H^\pm}]$. This conclusion
also applies to constraints derived from $R_{\ell 23}$ using
calculations of $f_K/f_\pi$ with a larger error (see section 2.2).
Hence it is clear that strict constraints on the plane $[\tan\beta,m_{H^\pm}]$ from 
$\rm{BR}(D^\pm_s \to \tau^\pm \nu_\tau/\mu^\pm \nu_\mu)$ and $R_{\ell 23}$
are strongly dependent on input from lattice QCD. However, their
tree-level sensitivity to $H^\pm$ makes them potentially valuable probes of the MSSM and
2HDM (Type~II), despite the above reservations.
Both $\rm{BR}(D^\pm_s \to \tau^\pm \nu_\tau)$ and $\rm{BR}(D^\pm_s \to \mu^\pm \nu_\mu)$
could start to play a significant role in constraining the plane $[\tan\beta,m_{H^\pm}]$
if future lattice calculations of $f_{D_s}$ favour $f_{D_s}<250$ MeV.
Given the ongoing interest in calculating $f_{D_s}$ in lattice QCD (summarized in
\cite{Gamiz:2008iv}) and the timely commencement of the BES-III experiment, 
we emphasize that these decays should not be overlooked in phenomenological
studies of $H^\pm$ in the MSSM and 2HDM (Type~II).

\begin{figure}[!t]
\begin{center}
\includegraphics[width=9.0cm]{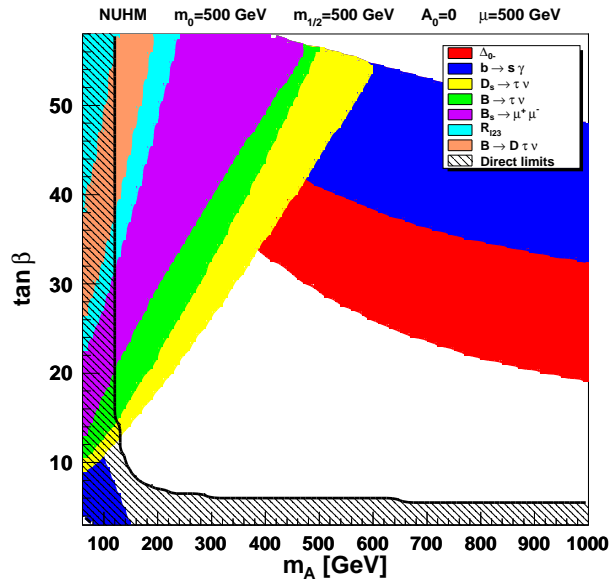}
\caption{Regions of the plane $[\tan\beta,m_A]$ in the NUHM model excluded by different flavour observables. 
The constraints are superimposed in the order given in the legend. }
\label{combined_fig}
\end{center}
\end{figure}

Finally we discuss the constraints on the plane $[\tan\beta,m_{H^\pm}]$ from $B$ physics observables.
As discussed in the introduction, one-loop induced decays such as 
$b\to s\gamma$ \cite{Bertolini:1990if} and $B^0_s\to \mu^+\mu^-$
\cite{Choudhury:1998ze,Babu:1999hn}
can give powerful constraints on the plane $[\tan\beta,m_{H^\pm}]$ (e.g. \cite{Isidori:2006pk,Eriksson:2008cx,Carena:2006ai,Ellis:2007fu,Mahmoudi:2007gd})
although with a stronger dependence on 
SUSY parameters than the purely leptonic decays discussed up to now.
In Fig.~\ref{combined_fig} we present an example of the constraints obtained by different flavour observables 
(both tree-level and one-loop induced) for 
fixed values of $m_0 = m_{1/2} = \mu = 500$ GeV and $A_0 = 0$. We show the results in the plane $[\tan\beta,m_A]$,
where $m_A$ is the mass of the pseudoscalar which is very close in magnitude to $m_{H^\pm}$ for
$m_A > 200$ GeV. The regions excluded by $b \to s \gamma$ observables are 
displayed in red for the isospin asymmetry and in blue for the branching ratio \cite{Mahmoudi:2007gd,Ahmady:2006yr}. 
The region excluded by $\rm{BR}(D^\pm_s \to \tau^\pm \nu)$ is depicted in yellow. The green area represents the region excluded
 by $\rm{BR}(B^\pm \to \tau^\pm \nu)$, the violet region by $\rm{BR}(B^0_s \to \mu^+ \mu^-)$, 
the light blue region by $K^\pm \to \mu^\pm \nu$, and the orange area by 
$\rm{BR}(B \to D \tau \nu)$ \cite{Eriksson:2008cx}, the latter also being sensitive at tree level to $H^\pm$ and
can provide constraints competitive with $\rm{BR}(B^\pm \to \tau^\pm \nu)$ \cite{Itoh:2004ye,Chen:2006nua,Nierste:2008qe,Grzadkowski:1991kb,Kamenik:2008tj}.
To obtain the constraints presented in this figure the input values of \cite{Mahmoudi:2008tp} are used. It is important to 
remember that the constraints can be subject to uncertainties, in particular from decay constants and CKM matrix elements. 
To obtain the constraint from $\rm{BR}(D^\pm_s \to \tau^\pm \nu)$ the central value $m_s/m_c=0.08$ is used. Finally, the black 
region in the figure represents the region excluded by the direct searches at colliders \cite{Amsler:2008zzb}.
This figure shows that $\rm{BR}(D^\pm_s \to \tau^\pm \nu)$ can be very competitive with the other decays in the context of the NUHM.

\section{Conclusions}
Constraints on the parameter space of $[\tan\beta,m_{H^\pm}]$ have been derived from the
effect of singly charged Higgs bosons on the decays $D^\pm_s\to \mu^\pm\nu_{\mu}$ and 
$D^\pm_s\to \tau^\pm\nu_{\tau}$ in the context of the Minimal Supersymmetric Standard Model.
It was shown that such constraints can be competitive with and complementary to those derived from 
other flavour physics observables, especially if future lattice QCD calculations favour $f_{D_s}< 250$ MeV 
(which is already suggested by the result of \cite{Follana:2007uv}).
We emphasize that such decays will be measured more precisely at the charm facility
BES-III and hence could play an important role in constraining the plane
$[\tan\beta,m_{H^\pm}]$ in the context of supersymmetric models in the future.
The numerical analysis was performed by the publically available code
SuperIso \cite{Mahmoudi:2007vz,Mahmoudi:2008tp} and we encourage the inclusion of 
$D^\pm_s\to \mu^\pm\nu_{\mu}$ and  $D^\pm_s\to \tau^\pm\nu_{\tau}$ in other studies of flavour physics constraints on supersymmetric models.

\section*{Acknowledgements}
AGA was supported by the ``National Central University Plan to Develop First-class Universities 
and Top-level Research Centres''. Comments from B. Sciascia are gratefully acknowledged.

\bibliographystyle{JHEP}
\bibliography{Ds_paper}

\end{document}